\documentclass[11pt]{article}
\usepackage{amsmath,amsfonts,amssymb,bbm}
\usepackage{tensor}
\usepackage[T1]{fontenc}
\usepackage[utf8]{inputenc}
\usepackage{lmodern}
\usepackage{authblk}
\usepackage[disable]{todonotes}
\usepackage{color}
\usepackage{multicol}
\usepackage{cite}
\usepackage{collref}
\usepackage[margin=3cm]{geometry}
\usepackage{hyperref}

\newcommand*\xbar[1]{%
  \hbox{%
    \vbox{%
      \hrule height 0.5pt % The actual bar
      \kern0.3ex%         % Distance between bar and symbol
      \hbox{%
        \kern-0.0em%      % Shortening on the left side
        \ensuremath{#1}%
        \kern-0.0em%      % Shortening on the right side
      }%
    }%
  }%
}

\newcommand\Preind[3]{\vphantom{#3}#1#2#3}

\newcommand\RR{\ensuremath{\mathbb R}}

\newcommand\email[1]{\thanks{\href{mailto:#1}{\nolinkurl{#1}}}}
\newcommand\td{\text{d}}

\newcommand\cO{{\cal O}}

\newcommand{\skyp}{{\cal I}^+}
\newcommand{\skym}{{\cal I}^-}
\newcommand{\bz}{\bar{z}}
\newcommand{\bw}{\bar{w}}

\newcommand{\p}{\partial}
\newcommand{\pb}{\p_{\bz}}
\newcommand{\be}{\begin{equation}}
\newcommand{\ee}{\end{equation}}
\newcommand{\bea}{\begin{eqnarray}}
\newcommand{\eea}{\end{eqnarray}}

\newcommand{\half}{\frac{1}{2}} %%% WTF Someone explains to me why \half is faster than \frac12 ;P
\def \ga {\gamma_{z\bz}}
\def \gai {\gamma_{z\bz}^{-1}}
\def \sg {\sqrt{-g}}

\def \bw {\bar{w}}
\def \gawi {\gamma_{w\bw}^{-1}}


\author[a]{Eduardo Conde\email{econdepe@snu.ac.kr}\,}
\author[b]{Pujian Mao\email{pujian.mao@ulb.ac.be}\,}

\affil[a]{\,School of Physics \& Astronomy and Center for Theoretical Physics, Seoul National University, Seoul 08826, South Korea}
\affil[b]{\,Service de Physique Th\'eorique et Math\'ematique, Universit\'e Libre de Bruxelles and International Solvay Institutes, Campus de la Plaine, CP 231, B-1050 Bruxelles, Belgium}
\title{\huge\bf Remarks on Asymptotic Symmetries and the Sub-leading Soft Photon Theorem\\}
\date{}

\begin{document}
 \maketitle
 \thispagestyle{empty}

\begin{abstract}
A deep connection has been recently established between soft theorems and symmetries at null infinity in gravity and gauge theories, recasting the former as Ward identities of the latter.
In particular, different orders (in the frequency of the soft particle) in the soft theorems are believed to be controlled by different asymptotic symmetries.
In this paper we argue that this needs not be the case by focusing on the soft photon theorem.
We argue that the sub-leading soft factor follows from the same symmetry responsible for the leading one, namely certain residual (large) gauge transformations of the gauge theory.
In particular, expanding the associated charge in inverse powers of the radial coordinate, the (sub-)leading charge yields the (sub-)leading soft factor.
\end{abstract}

\cleardoublepage

\section{Introduction}

Many ordinary flat-space Quantum Field Theories without mass gap, like gravity or perturbative gauge theories plus their different supersymmetric versions, have revealed to possess simpler S-matrices than expected. One may wonder if this could be somehow connected to the particular structure of null infinity of asymptotically flat space-times, which is accessible to the massless excitations of the theory. A connection along these lines has been recently suggested by Strominger~\cite{Strominger:2013lka,Strominger:2013jfa}, who proposed to identify soft theorems with Ward identities of asymptotic symmetries, building a bridge between two previously unrelated subjects.

This bridge has attracted a lot of attention recently, boosting new promising lines of research, like possible attacks on the information paradox~\cite{Hawking:2016msc} or holographic ideas for flat space~\cite{Kapec:2016jld,Cheung:2016iub}. A key role in exploiting this bridge is the identification of the symmetries at null infinity of the scattering. Indeed, for the gravitational scattering, the knowledge of a large family of asymptotic symmetries~\cite{Bondi:1962px,Sachs:1962wk,Barnich:2009se,Barnich:2010eb,Barnich:2011ct,Barnich:2011mi} prompted Cachazo and Strominger to conjecture new soft graviton theorems~\cite{Cachazo:2014fwa} (see \cite{Gross:1968in,White:2011yy} for earlier, and less general, versions of this theorem). These new soft theorems\footnote{For gluons also a new soft theorem has been conjectured in~\cite{Casali:2014xpa} (see~\cite{Laenen:2008gt,Laenen:2010uz} for earlier partial versions, completed in~\cite{White:2014qia}), although in that case it was not motivated by a known asymptotic symmetry.} come at sub-leading orders in the frequency of the soft particle. Conversely, one could presumably use known soft theorems to uncover novel symmetries at null infinity. However, in this paper we critically argue that this is not a one-to-one correspondence.

More precisely we show that sub-leading orders in soft theorems are not necessarily always associated to new asymptotic symmetries. These sub-leading soft theorems may be simply associated with the structure of (previously known) symmetries \textit{near} null infinity. While we illustrate this idea with the soft photon theorem, resolving a previous puzzle in the literature, the technique we present is of wider applicability.

Our plan is the following. We start by explaining the current status of the connection between the soft photon theorem and asymptotic symmetries, where the leading soft photon factor has been linked to residual (large) gauge transformations. Afterwards we rigorously analyze the asymptotic structure of the classical solutions of the Maxwell system coupled to a general source. This allows us to formally expand in powers of $1/r$ the charge that can be associated to residual gauge transformations. Finally we show how the second piece in this expansion gives us the sub-leading soft photon theorem. After some final remarks, we include a somewhat tangential appendix on the Newman-Penrose charges, expressing them in a language more akin to the one used throughout the rest of the paper.

\section{Soft Photon Theorem}

The origins of the soft photon theorem go back to the 50's, when Low~\cite{Low:1954kd} and Gell-Mann and Goldberger~\cite{GellMann:1954kc} realized that the scattering of light by arbitrary targets displayed universal properties through the next-to-leading order in a low-frequency expansion. Such a result was put in amplitude form by Low~\cite{Low:1958sn}:
\begin{equation}
\label{subsofth}
	M_{n+1}=\left(S_{\pm}^{(0)}+S_{\pm}^{(1)}\right)M_{n}+\cO\left(\omega\right) \  ,
\end{equation}
where the extra particle on the left-hand-side amplitude is a soft photon of helicity $\pm1$, and frequency $\omega$. On the right-hand side we find the so-called leading and sub-leading (in $\omega$) soft factors:
\begin{equation}
\label{Ss}
  S_{\pm}^{(0)}=\sum_{k=1}^n e_k\,\frac{p_k\cdot \epsilon^{\pm}}{p_k\cdot q} \ ,\quad
  S_{\pm}^{(1)}=\sum_{k=1}^n e_k\,\frac{q_{\mu}\epsilon_{\nu}^{\pm}\,J_k^{\mu\nu}}{p_k\cdot q} \ .
\end{equation}
These involve the momentum of the soft photon, $q$, its polarization tensor $\epsilon^{\pm}$, and the momenta and angular momentum operators of the rest of the particles $k=1,\ldots,n$ in the reaction. The tree-level relation~\eqref{subsofth} is universal in the sense that it holds for any type of matter minimally coupled\footnote{
Soft theorems can be modified in the presence of non-minimal couplings, as it happens for instance for the sub-sub-leading tree-level soft graviton theorem~\cite{Bianchi:2014gla}.
}
to Maxwell theory~\cite{Burnett:1967km,Bell:1969yw}.

More excitingly, the leading soft factor $S_{\pm}^{(0)}$ does not get loop corrected~\cite{Weinberg:1965nx}.\footnote{
Instead loop corrections do affect the form of the sub-leading soft factor $S_{\pm}^{(1)}$~ \cite{DelDuca:1990gz,Larkoski:2014bxa}. These corrections are beyond the IR divergencies associated to massless sources, which can be taken care of by the Kulish-Faddeev formalism~\cite{Kulish:1970ut} (see~\cite{Gomez:2016hxz} for a recent extension of this formalism to sub-leading order).
}
This may suggest the existence of some symmetries operating at the quantum level. Following Strominger, soft theorems should be the Ward identities associated to asymptotic symmetries. To make a more precise identification, we need to pinpoint which are the symmetries responsible for the different terms in a soft theorem, and their corresponding charges.

In \cite{He:2014cra} (see also~\cite{Kapec:2014zla,Mohd:2014oja,Kapec:2015ena,Campiglia:2015qka,Strominger:2015bla}) it was found that certain residual (large) gauge transformations are responsible for the leading piece in~\eqref{subsofth}. Regarding the sub-leading order, the authors in~\cite{Lysov:2014csa} derived from the known sub-leading soft factor~\eqref{Ss} the form that the charge of the asymptotic symmetry should have, although the origin and nature of the symmetry could not be identified. We address this issue in this paper.

If we search for inspiration in the old works on the soft photon theorem \cite{Low:1954kd,GellMann:1954kc,Low:1958sn,Burnett:1967km,Bell:1969yw}, or also more recent ones~\cite{Broedel:2014fsa,Bern:2014vva}, we notice that the fundamental ingredient to explain both terms in~\eqref{subsofth} was only gauge invariance. That may lead one to think that \textit{no new symmetry} needs to be invoked. Or in other words, that the residual large gauge transformations responsible for the leading soft factor can also explain the sub-leading one. This is precisely what we show in the next pages.

\section{Solution Space For Four-Dimensional Maxwell Theory}
\label{sos}

The solution space of the source-free Maxwell equations has been discussed in \cite{Janis:1965tx,Newman:1968uj} in the Newman-Penrose formalism. Contrary to the approach these old works took, here we favor the use of the gauge field $A_\mu$ over the field strength $F_{\mu\nu}$.
Our analysis is similar in spirit to that in \cite{Barnich:2015jua}, where the asymptotic structure at null infinity of three-dimensional Einstein-Maxwell theory was worked out.

Our theory lives in Minkowski space-time, which has two null ``boundaries'', past null infinity $\skym$, and future null infinity $\skyp$. They are better appreciated when adopting advanced or retarded coordinates respectively. In what follows we concentrate on $\skyp$, although everything can be similarly repeated on $\skym$. We introduce retarded spherical coordinates with the following change of coordinates:
\begin{equation}
\label{retard}
u=t-r\,,\quad x^1+ix^2=\frac{2rz}{1+z\bz}\,,\quad x^3=r\,\frac{1-z\bz}{1+z\bz}\,.
\end{equation}
where $r=\sqrt{x^ix_i}$. Minkowski space-time becomes
\begin{equation}
\label{metric}
\td s^2=-\td u^2-2\td u\td r+2r^2\gamma_{z\bz}\td z\td\bz\,,\quad \gamma_{z\bz}=\frac{2}{(1+z\bz)^2}\,.
\end{equation}
The piece $2r^2\gamma_{z\bz}\td z\td\bz$ is just the metric of the round sphere $S^2$, and $\skyp$ is precisely the submanifold $r=\infty$ in the retarded spherical coordinates, with topology $S^2\times\RR$. The sphere at $u=\pm\infty$ is denoted by $\skyp_\pm$. In the bulk we have a gauge field $A_{\mu}$ and a conserved matter current $J_{\mu}$. As \cite{He:2014cra}, we choose the following (radial) gauge and asymptotic conditions:
\begin{equation}
\label{asycond}
  \begin{aligned}
    A_r&=0 \ , & A_u&=\cO(r^{-1}) \ , & A_z&=\cO(1) \ ,\\
    J_r&=0 \ , & J_u&=\cO(r^{-2}) \ , & J_z&=\cO(r^{-2}) \ .
  \end{aligned}
\end{equation}
We have used the ambiguities of a conserved current\footnote{When a conserved current is derived from a global symmetry, it is naturally defined up to the equivalence $J^\mu\thicksim J^\mu+\nabla_\nu k^{[\mu\nu]}$, so it makes more sense to consider equivalence classes of currents $[J^\mu]$ (see \cite{Barnich:2001jy} for details).} to set the radial component of the current to zero. This is consistent with working in the radial gauge.
More specifically, let us assume the following ansatz for the $\frac{1}{r}$-expansion of the gauge field
\begin{equation}
\label{Azz}
  \begin{aligned}
	  A_u&=\frac{A^0_u(u,z,\bz)}{r}+\cO\Big(\frac{1}{r^2}\Big) \ ,\\
	  A_{z(\bz)}&=A^0_{z(\bz)}(u,z,\bz) + \sum\limits_{m=1}^\infty \frac{A^m_{z(\bz)}(u,z,\bz)}{r^m} \ ,
	\end{aligned}
\end{equation}
and the current
\begin{equation}
  \begin{aligned}
  	J_u&=\frac{J^0_u(u,z,\bz)}{r^2}+\cO\left(\frac{1}{r^3}\right) \ , \\
  	J_{z(\bz)}&=\frac{J^0_{z(\bz)}(u,z,\bz)}{r^2} + \sum\limits_{m=1}^\infty \frac{J^m_{z(\bz)}(u,z,\bz)}{r^{m+2}} \ .
  \end{aligned}
\end{equation}
The reason for not specifying further the expansions of the $u$-components of $A_{\mu}$ and $J_{\mu}$ is that they are determined by the equations of motion. Indeed, we can solve the Maxwell equations, $\nabla_\mu F^{\mu \nu}=J^{\nu}$, and integrate the current conservation condition, $\nabla_\mu J^\mu=0$, in a $1/r$-expansion. We obtain:
\begin{align}
\label{current}
  &J_u=\frac{J^0_u(u,z,\bz)}{r^2} - \frac{1}{r^2} \int_r^{\infty}\td r'\left[ \gai(\p_z J_{\bz} + \p_{\bz} J_z)\right] \ ,\\
\label{Au}
  &A_u=\frac{A^0_u(u,z,\bz)}{r} +\nonumber\\
  &\int_{r}^{\infty} \td r'\, \frac{1}{r'^2} \int_{r'}^{\infty}\td r''\left[ \gai(\p_z \p_{r''} A_{\bz} + \p_{\bz}\p_{r''} A_z)\right] \ ,\\
\label{supeq}
  &\p_u A^0_u  =\gai \p_u(\p_z A^0_{\bz} + \p_{\bz} A^0_z) + J^0_u \ ,\\
\label{puA1}
  &2\p_u A_z^1=\p_z A^0_u  + \p_z [\gai(\p_z A^0_{\bz} - \p_{\bz} A^0_z)] + J^0_z \ ,\\
\label{puAm}
  &\p_u A_z^{m} = \frac{J^{m-1}_z}{2m} -\frac{m-1}{2}A_z^{m-1} - \frac{\p_z [\gai(\p_{\bz} A^{m-1}_z)]}{m} \ ,
\end{align}
where $m\geq2$ on the last line. We have omitted two extra equations, that can be obtained from the last two above swapping $z\rightleftharpoons\bz$. We notice that all the coefficients in the expansion of $A_{z(\bz)}$ have uniquely determined retarded time derivatives except the leading $A^0_{z(\bz)}$. From here on, we refer to $\partial_u A^0_z$ as the electromagnetic ``\textit{news}'' since it reflects the propagation of electromagnetic waves.

To summarize, the general solution to the Maxwell system in four-dimensional Minkowski space-time~\eqref{metric} with the prescribed asymptotics~\eqref{asycond} is completely determined in terms of the initial data $A^0_u(u_0,z,\bz)$, $A^m_z(u_0,z,\bz),A^m_{\bz}(u_0,z,\bz)$ ($m\geq1$), the functions $A^0_z(u,z,\bz)$, $A^0_{\bz}(u,z,\bz)$ and the matter current. The latter is characterized by the source functions $J^0_u(u,z,\bz)$, $J^m_z(u,z,\bz),J^m_{\bz}(u,z,\bz)$ ($m\geq0$).

\section{Charges}
\label{sec:charge}

The radial gauge condition that we imposed in~\eqref{asycond} leaves residual gauge transformations of the form $\delta A_{\mu}=\partial_{\mu}\varepsilon(z,\bz)$, where $\varepsilon(z,\bz)$ is an arbitrary function on the sphere. These transformations can be interpreted as symmetries at null infinity,
\begin{equation}
\label{large.g}
  \delta A^0_{z}=\partial_{z}\varepsilon(z,\bz) \ ,\qquad\delta A^0_{\bz}=\partial_{\bz}\varepsilon(z,\bz) \ ,
\end{equation}
or symmetries of the S-matrix following \cite{Strominger:2013lka,He:2014cra}. The associated charge at future null infinity is \cite{Ashtekar:1987tt,Barnich:2001jy,He:2014cra}:
\begin{equation}
\label{charge}
  \begin{aligned}
    Q_{\varepsilon_\textrm{out}}&= \int_{\skyp_{-}} \td z \td\bz \,\ga\, \varepsilon(z,\bz)\, r^2\,F_{ru} \\
    &= -\int_{\skyp} \td z \td\bz \td u\,\ga\, \varepsilon(z,\bz)\, r^2\,\p_u F_{ru} \ ,
  \end{aligned}
\end{equation}
where in the second equality we assumed $A^0_u|_{\skyp_+}=0$, meaning that in the far future the system contains no bulk electric charge. Let us plug the expansion~\eqref{Azz} in~\eqref{Au}, yielding:
\be
\label{Fru.exp}
    r^2\p_u F_{ru}=-\partial_u A^0_u + \frac{\gai\p_u(\p_z A^1_{\bz} + \p_{\bz} A^1_z)}{r}
    + \sum\limits_{m=1}^\infty \frac{\gai \p_u(\p_z A^{m+1}_{\bz} + \p_{\bz} A^{m+1}_z)}{r^{m+1}} \ ,
\ee
This gives a $\frac{1}{r}$-expansion of the charge~\eqref{charge}:
\begin{equation}
\label{Cexp}
  Q_{\varepsilon_\textrm{out}}=Q^{(0)}_{\varepsilon_\textrm{out}}+\frac{Q^{(1)}_{\varepsilon_\textrm{out}}}{r}+\sum_{m=2}^{\infty}\frac{Q^{(m)}_{\varepsilon_\textrm{out}}}{r^{m}} \ .
\end{equation}
This is a formal expansion since the integration in~\eqref{charge} is performed at $r=\infty$. The fact that we are allowed to expand in $1/r$ quantities defined at infinity should be understood as the counterpart of the frequency expansion in~\eqref{subsofth}, where the amplitudes are strictly only defined for $\omega=0$. Then it is reasonable to think that sub-leading orders in $1/r$ may be in correspondence with sub-leading orders in $\omega$. Below we show that the sub-leading term in~\eqref{Cexp} has this physical interpretation.

Using~\eqref{supeq}, we can see that the leading piece,
\begin{equation}
\label{Q0}
  Q^{(0)}_{\varepsilon_\textrm{out}}=-\int_{\skyp_{-}} \td z \td\bz \,\ga\, \varepsilon(z,\bz)\, A_u^0 \ ,
\end{equation}
gives the charge that was identified in \cite{He:2014cra} as responsible for the leading soft photon theorem. We want to investigate here the sub-leading piece $Q^{(1)}_{\varepsilon_\textrm{out}}$. Using~\eqref{puA1} we can immediately write
\begin{equation}
  Q^{(1)}_{\varepsilon_\textrm{out}}=\int_{\skyp} \td z \td\bz \td u\, \varepsilon\,\p_u\left(\p_z A^1_{\bz} + \p_{\bz} A^1_z\right) \ .
\end{equation}
In order to have everything in terms of ``news'', we can massage the integral to obtain:
\begin{multline}
\label{Q1}
  Q^{(1)}_{\varepsilon_\textrm{out}}\!=\!\int_{\skyp}\!\! \td z \td\bz \td u\!
  \left[ u \left(D^2_zD^z\varepsilon\,\p_u A^0_{\bz} + D^2_{\bz}D^{\bz}\varepsilon\,\p_u A^0_z\right)-\right.\\
  \left.-\half u\left(D_z D_{\bz}\varepsilon + D_{\bz} D_z \varepsilon\right) J^0_u
  -\half\left(D_z \varepsilon\, J^0_{\bz} + D_{\bz}\varepsilon \, J^0_z\right)\right] \ ,
\end{multline}
where we have used again~\eqref{supeq}, and one has to drop the boundary term $\int_{\skyp} \td z \td\bz \td u\,\p_u\left(u\, \varepsilon\, \p_z\pb A^0_u\right)=-\lim_{u\to\infty} u\, \int_{\skyp_-} \td z \td\bz \, \p_z\pb\varepsilon\, A^0_u\,$. This second expression is zero when bracketed between \textit{in} and \textit{out} states\footnote{The $u$ factor in front should not be worrisome, as it must still be combined with a $1/r$ factor as in~\eqref{Cexp}.}, which is simply a consequence of $\langle\textrm{out}|Q^{(0)}_{\varepsilon_\textrm{out}}|\textrm{in}\rangle=0$. The notation $D_{z(\bz)}$ is for the two-dimensional covariant derivative on the sphere. If one now calls
\begin{equation}
  Y^z=D^z\varepsilon \ ,\qquad Y^{\bz}=D^{\bz}\varepsilon \ ,
\end{equation}
we see that the sub-leading charge $Q^{(1)}_{\varepsilon_\textrm{out}}$ is exactly the one that was written in~\cite{Lysov:2014csa}. This is the charge that one gets when translating the sub-leading soft photon theorem into an S-matrix Ward identity.%, as we review below.

\section{Soft Theorems}
\label{amp}

In order to better appreciate the interplay between leading and sub-leading terms in the present context, we give an overview of the results in \cite{He:2014cra} and \cite{Lysov:2014csa} adapted to the notation of this paper.

Let us start by rewriting the soft photon theorem~\eqref{subsofth} in the retarded spherical coordinates of~\eqref{retard}. We can characterize the soft momentum $q$ by an energy $\omega_q$ and a direction on the sphere, $w$, as
\begin{equation}
\label{qmu}
  q_{\mu}=\omega_q\left(1,\frac{w+\bar{w}}{1+w\bar{w}},i\frac{\bar{w}-w}{1+w\bar{w}},\frac{1-w\bar{w}}{1+w\bar{w}}\right)\ .
\end{equation}
For simplicity and following \cite{Lysov:2014csa}, let us assume massless scalar matter. Then the $n$ momenta in the amplitudes of~\eqref{subsofth} can be charaterized as~\eqref{qmu} by energies $\omega_k$ and directions $w_k$. The leading and sub-leading pieces of the soft photon theorem~\eqref{subsofth} are then rewritten as
\begin{gather}
\label{soft0+}
	\lim_{\omega_q\to0}\langle\textrm{out}|\omega_q\,\mathfrak{a}_+(q)|\textrm{in}\rangle=
	\frac{1+|w|^2}{\sqrt{2}}\sum_{k=1}^n \frac{e_k}{w-w_k}\,\langle\textrm{out}|\textrm{in}\rangle\ ,\\
\label{soft1+}
  \lim_{\omega_q\to0}\!\partial_{\omega_q}\langle\textrm{out}|\omega_q\,\mathfrak{a}_-(q)|\textrm{in}\rangle\!=\!
  \sum_{k=1}^n\! \frac{e_k}{\sqrt{2}}\!\left(\frac{1+w\,\bw_k}{\bw-\bw_k}\partial_{\omega_k}+
  \frac{(1+|w_k|^2)(w-w_k)}{\omega_k(\bw-\bw_k)}\partial_{w_k}\right)\!\langle\textrm{out}|\textrm{in}\rangle \ ,
\end{gather}
where $e_k$ is the electric charge of the $k$-th particle, and $\mathfrak{a}_{+(-)}(q)$ the annihilation operator which creates outgoing negative(positive)-helicity soft photons with momentum $q$. We have particularized to an outgoing negative(positive)-helicity soft photon for the leading (sub-leading) soft theorem, for reasons that become apparent below. Other cases can be treated in a similar manner.

To connect with asymptotic symmetries, we recall that in S-matrix language, a symmetry is just a relation between matrix elements
$\langle\rm{out'}|\rm{in'}\rangle=\langle\rm{out}|\rm{in}\rangle$, where the \textit{in} and \textit{out} states are transformed as
$|\rm{in'/out'}\rangle=U^{\textrm{in/out}}|\rm{in/out}\rangle$. The operators implementing the symmetry must verify ${U^{\textrm{out}}}^\dagger U^{\textrm{in}}=1$. If this symmetry is generated by a charge $Q$, the associated Ward identity reads as
\begin{equation}
\label{Ward}
	\langle\rm{out}|Q^{\textrm{out}}-Q^{\textrm{in}}|\rm{in}\rangle=0 \ .
\end{equation}
The charge for a spontaneously broken symmetry must act non-linearly on the states -- otherwise it would annihilate the vacuum -- so it can be decomposed into linear and non-linear pieces $Q=Q_{\rm{L}}+Q_{\rm{NL}}$. The Ward identity for a broken charge becomes
\begin{equation}
\label{WardNL}
	\langle\rm{out}|Q_{\rm{NL}}^{\textrm{out}}-Q^{\textrm{in}}_{\rm{NL}}|\rm{in}\rangle=
	-\langle\rm{out}|Q^{\textrm{out}}_{\rm{L}}-Q^{\textrm{in}}_{\rm{L}}|\rm{in}\rangle\,.
\end{equation}
Neglecting issues about a proper, non-divergent definition of a broken charge, if $Q_{\textrm{NL}}$ creates zero-momentum Goldstone bosons we can appreciate the similitude between equations~\eqref{WardNL} and~\eqref{subsofth}.

Let us now decompose the leading and sub-leading charges of~\eqref{Q0},~\eqref{Q1} into a piece containing the ``news'' and the rest, containing the sources. These are respectively the non-linear and linear pieces above. For notational brevity, we momentarily suppress the \textit{out} label and keep only the anti-holomorphic terms. The leading charge splits as
\begin{align}
\label{Q0NL}
	Q^{(0)}_{\textrm{NL}}=\int_{\skyp}\td z\td\bz \td u\,\varepsilon\,\partial_u\partial_{\bz}A^0_z \ ,\\
\label{Q0L}
	Q^{(0)}_{\textrm{L}}=\frac12\int_{\skyp}\td z\td\bz \td u\,\gamma_{z\bz}\,\varepsilon\,J_u^0 \ ,
\end{align}
while for the sub-leading we have
\begin{align}
\label{Q1NL}
	Q^{(1)}_{\textrm{NL}}&=\int_{\skyp} \td z \td\bz \td u\,u \left(D^2_zD^z\varepsilon\,\p_u A^0_{\bz}\right) \ ,\\
\label{Q1L}
	Q^{(1)}_{\textrm{L}}&=-\frac12\int_{\skyp} \td z \td\bz \td u\left(u\, D_z D_{\bz}\varepsilon\, J^0_u+
	D_{\bz}\varepsilon \, J^0_z\right) \ .
\end{align}
To see how~\eqref{soft0+} and~\eqref{soft1+} arise from~\eqref{WardNL}, one just needs to plug in there the (\textit{out}) charges~\eqref{Q0NL}-\eqref{Q1L} (plus their analogues for the \textit{in} charges, written in~\cite{He:2014cra,Lysov:2014csa}), and make a concrete choice of $\varepsilon(z,\bz)$. For the cases under consideration, the convenient choice is
\begin{equation}
\label{eps}
	\varepsilon(z,\bz)=\frac{1}{w-z} \ ,
\end{equation}
for which $\partial_{\bz}\varepsilon=-2\pi\delta^2(z-w)\,$. In order to obtain the proper action of the charges on the \textit{out} states, one has to define canonical commutation relations at infinity \cite{Ashtekar:1987tt}.  For the ``news'' fields, it is enough to perform a stationary-phase approximation of the gauge-field mode expansion:
\begin{equation}
	A^0_{z(\bz)}(x)=-\frac{i}{8\pi^2}\frac{\sqrt{2}}{1+z\bz}\int_0^\infty\td \omega_q\left[
\mathfrak{a}_{+(-)}(\omega_q\hat{x})\,e^{-i\omega_q u}-\mathfrak{a}^\dagger_{-(+)}(\omega_q\hat{x})\,e^{i\omega_q u}\right] \ ,
\end{equation}
where the creation and annihilation operators satisfy the standard commutation relations. Then, using the simple Fourier relations (defining $F(u)=\int_{-\infty}^{\infty}\td\omega\,e^{i\omega u}\tilde{F}(\omega)$):
%\begin{align}
%	\int\td u\,\partial_uF(u)&=\int\td\omega\td u\,\partial_ue^{i\omega u}\tilde{F}(\omega)=
%	2\pi i\lim_{\omega\to0}\left[\omega\tilde{F}(\omega)\right] \ , \\
%	\int\td u\,u\,\partial_uF(u)&=-2\pi \lim_{\omega\to0}\left[\partial_{\omega}\left(\omega\tilde{F}(\omega)\right)\right] \ ,
%\end{align}
\begin{equation}
	\int_{-\infty}^{\infty}\td u\,\partial_uF(u)=2\pi i\lim_{\omega\to0}\left[\omega\tilde{F}(\omega)\right] \ ,\quad
	\int_{-\infty}^{\infty}\td u\,u\,\partial_uF(u)=-2\pi \lim_{\omega\to0}\left[\partial_{\omega}\left(\omega\tilde{F}(\omega)\right)\right] \ ,
\end{equation}
and the special form that that $\partial_{\bz}\varepsilon=-2\pi\delta^2(z-w)$ takes for the choice~\eqref{eps}, we obtain for the non-linear pieces of the charges\footnote{
We are just keeping the anti-holomorphic parts of the the charges~\eqref{Q0NL}-\eqref{Q1L}, meaning those containing only $\partial_{\bz}\varepsilon$. In particular one has to split~\eqref{Q0L} via $J^0_u\to\frac12J^0_u+\frac12J^0_u$, which is essentially the same arbitrary separation we have for the first term in~\eqref{Q1L}. Otherwise one has to
introduce some extra factors of 2, that arise from a proper treatment of the radiative phase space~\cite{He:2014cra,Mohd:2014oja}.
}:
\begin{align}
\label{Q0NLbk}
	\langle\textrm{out}|Q^{(0)}_{\textrm{NL}}|\textrm{in}\rangle&=\frac{1}{4}\frac{\sqrt{2}}{1+|w|^2}\lim_{\omega_q\to0}
	\langle\textrm{out}|\omega_q\,\mathfrak{a}_+(q)|\textrm{in}\rangle \ , \\
\label{Q1NLbk}
  \langle\textrm{out}|Q^{(1)}_{\textrm{NL}}|\textrm{in}\rangle&=
  -\frac{\sqrt{2}i}{4\gamma_{w\bw}}\,D^2_w\left[\frac{1}{1+|w|^2}\lim_{\omega_q\to0}\!\partial_{\omega_q}
  \langle\textrm{out}|\omega_q\,\mathfrak{a}_-(q)|\textrm{in}\rangle\right] \ ,
\end{align}
where we are denoting $\gamma_{w\bw}=\frac{2}{(1+|w|^2)^2}$.
Regarding the linear pieces, restricting ourselves to scalar charged (with charge $Q_e$) matter: $\Phi=\sum_{m=0}^\infty\frac{\Phi^m(u,z,\bz)}{r^{m+1}}$ with current $J^0_\mu=i Q_e(\bar{\Phi}^0\p_\mu \Phi^0 - \Phi^0\p_\mu \bar{\Phi}^0)$ at leading order\footnote{Recall from~\eqref{asycond} we are taking the current to have zero radial component. This can be done by adding a total derivative to the usual current as $J^\mu=i Q_e(\bar{\Phi}\nabla^\mu \Phi - \Phi\nabla^\mu \bar{\Phi})+\nabla_\nu k^{[\mu\nu]}\,$, with an anti-symmetric two-tensor whose only non-zero component is $k^{[ur]}=\dfrac{1}{\sg}\int\td r\,\sg(\xbar\Phi\p_r\Phi-\Phi\p_r\xbar\Phi)\,$. Therefore $J^0_r=0\,$.}, we just need to use the boundary canonical commutation relation \cite{Lysov:2014csa}:
\begin{equation}
  [\bar{\Phi}^0(u,z,\bz),\Phi^0(u',w,\bw)]=\frac{i}{4}\gawi\, \Theta(u'-u)\delta^2(z-w) \ ,
\end{equation}
to see that
\begin{align}
\label{Q0Lbk}
  \langle\textrm{out}|Q^{(0)}_{\textrm{L}}|\textrm{in}\rangle&=\sum_{k=1}^n
  -\frac{e_k}{4(w-w_k)}\langle\textrm{out}|\textrm{in}\rangle \ ,\\
\label{Q1Lbk}
  \langle\textrm{out}|Q^{(1)}_{\textrm{L}}|\textrm{in}\rangle&=\sum_{k=1}^n
  -\frac{\pi i\,e_k}{2}\left(\partial_w(\gamma_{w\bw}^{-1}\delta(w-w_k))\partial_{\omega_k}+\gamma_{w\bw}^{-1}\,\omega_k^{-1}\delta(w-w_k)\partial_{w_k}\right)\langle\textrm{out}|\textrm{in}\rangle \ .
\end{align}
Assembling all these expressions, it is immediate to recover the leading soft theorem~\eqref{soft0+} from~\eqref{WardNL}. To also recover  the sub-leading~\eqref{soft1+}, we just need to use the extra identity:
\begin{multline}
\label{D2w}
D^2_{w}\left[\frac{1}{1+|w|^2}\left(\frac{1+w\,\bw_k}{\bw-\bw_k}\partial_{\omega_k}+
\frac{(1+|w_k|^2)(w-w_k)}{\omega_k(\bw-\bw_k)}\partial_{w_k}\right)\right]\\
=-2\pi\,\gamma_{w\bw}\left(\partial_w\left(\gamma_{w\bw}^{-1}\delta(w-w_k)\right)\partial_{\omega_k}
+\gamma_{w\bw}^{-1}\,\omega_k^{-1}\delta(w-w_k)\partial_{w_k}\right) \ .
\end{multline}
Let us notice that a resembling phenomenon happens in theories with spontaneously broken conformal symmetry, where dilatations (conformal boosts) control the (sub-)sub-leading soft dilaton theorem. The modes generated by dilatations are also given essentially by the derivative of the modes generated by conformal boost~\cite{DiVecchia:2015jaq}.

We recall that we have only paid attention to the \textit{out} part of~\eqref{WardNL}. The analysis of the \textit{in} part can be carried out analogously, up to an anti-podal identification. These details have been laid out in~\cite{He:2014cra,Lysov:2014csa}, and are not needed for our only-illustrative purposes of this section, which does not contain new results.

\section{Discussion}
\label{bsoq}

We have shown how the connection between the leading soft photon theorem and the asymptotic symmetry associated to residual large gauge transformations of the Maxwell system can be extended to the sub-leading order. By properly studying the asymptotic structure of Maxwell equations, we have seen that the charge associated to the asymptotic symmetry~\eqref{large.g} can be formally expanded in powers of $\frac{1}{r}$. This induces several Ward identities for such a charge. It was already established in~\cite{He:2014cra} that the $\cO(r^0)$ Ward identity was equivalent to the leading soft photon theorem. Our finding here is that the $\cO(r^{-1})$ Ward identity produces the sub-leading soft photon theorem. In particular, the $\cO(r^{-1})$ term of the charge~\eqref{charge} matches the one conjectured in~\cite{Lysov:2014csa}.

Since our reasoning is not particularly tied to the Maxwell system, we expect it to be a general feature of the connection between soft theorems and asymptotic symmetries. Namely, certain asymptotic symmetries will give rise to several soft theorems via the mechanism we illustrated in this paper, where sub-leading (in frequency) soft factors will come from the sub-leading (in $1/r$) behavior of asymptotic symmetries.

An obvious extension of this work is to apply the technique to the case of pure gravity, where the tree-level soft theorem stretches to sub-sub-leading order. While certain coincidences make that case a bit deceiving, with supertranslations and superrotations seemingly controlling leading and sub-leading orders respectively~\cite{Kapec:2014opa,Campiglia:2014yka,Campiglia:2015yka}, we expect these two symmetries to be sufficient to explain the sub-sub-leading order~\cite{2come}.

%The charge~\eqref{charge} actually contains an infinite number of charges, parametrized by the function $\varepsilon(z,\bz)$. A curious observation is that the function that naturally gives the form of the leading soft theorem for an outgoing photon with negative helicity, namely~\eqref{eps}, gives the sub-leading soft theorem for an outgoing photon with positive helicity (differentiated twice with respect to the angular direction of the soft photon).

Another question worth commenting is the following. In light of the procedure presented here, it seems that one should get an infinite number of soft theorems, by just considering the remaining $\cO(r^{-n})$ ($n\geq2$) orders of the charge. We can see from~\eqref{Fru.exp} and~\eqref{puAm} that this is not the case. There is no way to massage these sub-sub-leading charges the way we did in~\eqref{Q1} without running into infinities.

A more intuitive way to understand this is to notice that the boundary fields, defined as those appearing at leading order in the different components of the field-strength~\cite{He:2014cra}, are only $A^0_u$, $A^0_{z(\bz)}$ and $A^1_{z(\bz)}$. Only the charges that involve these fields are physically meaningful, since the charges are defined at infinity. Therefore, as soon as $A_z^m$ ($m\geq2$) appears in the charge, which happens at sub-sub-leading order, we do not obtain new soft theorems. In hindsight, looking back for instance at the relevant Feynman diagrams in~\cite{Bern:2014vva}, this phenomenon can be seen as the counterpart of the fact that gauge invariance does not fix beyond the sub-leading order photon emission from internal legs.

Finally, with the semi-classical approach we have adopted in this work, we do not expect to see the loop corrections that should appear at sub-leading order in the soft photon theorem. However, we note here that in our shift of the current $J_{\mu}$ we have actually set to zero some integration constants. While the leading charge~\eqref{Q0} is unaffected by these constants, the sub-leading charge~\eqref{Q1} is sensitive to them. It might be possible that keeping track of new quantum-generated interactions, one could adapt the formalism presented in this manuscript to account for such loop corrections. We leave that line open for future investigations.

\section*{Acknowledgments}
We would like to thank Glenn Barnich, Yu-tin Huang, Euihun Joung, Jiang Long, Soo-jong Rey and Hongbao Zhang for useful discussions on these and related matters, and additionally Yu-tin Huang again for valuable comments on the draft.
The work of EC was supported in part by the National Research Foundation of Korea through the grant NRF-2014R1A6A3A04056670, and the grants 2005-0093843, 2010-220-C00003 and 2012K2A1A9055280. P.~Mao is supported by a PhD fellowship from the China Scholarship Council.

\appendix

\section{Newman-Penrose charges}

As proposed in~\cite{He:2014cra}, it is very suggesting to think of the charge controlling the leading soft photon theorem as a generalization of the usual electric charge, so as to make it dependent on the angle at the $S^2$ at infinity.
In the late 60's Newman and Penrose~\cite{Newman:1968uj} discovered new conserved charges for several theories possessing massless particles. These charges cannot be associated to bulk divergenceless vectors as usual; instead they are expressed as surface integrals at infinity.
For the Maxwell theory, of course one of these Newman-Penrose charges is the electric charge.

It would seem reasonable to think that the charge in control of the sub-leading soft photon theorem should be an angle-dependent generalization of the other Newman-Penrose charges. It was hinted in~\cite{Lysov:2014csa} that one should consider generalizing the ``sub-leading'' charge in a multipole expansion of the electromagnetic field, namely dipole charge. In order to investigate this possibility,
we will review the main results of the Maxwell theory part in~\cite{Newman:1968uj} in this appendix.

First we collect some mathematical results. We define the $\eth,\xbar\eth$ operators as
\begin{equation}
	\eth \eta=\ga^{-\half}\,\p_{\bz} \eta + s\,\eta\,\p_{\bz} \ga^{-\half} \ ,\qquad
	\xbar \eth \eta=\ga^{-\half}\,\p_z \eta - s\, \eta\,\p_{z} \ga^{-\half} \ ,
\end{equation}
where $s$ is the spin weight of the field $\eta$, meaning that it has the commutation relation\footnote{There is no factor $2$ compared to \cite{Newman:1968uj} since we are working on a unit sphere.} $[\eth,\xbar\eth]\eta=-s\eta$. An important property of $\eth$ and $\xbar\eth$ is their action on the spherical harmonics $Y_{l,m}$ $(l=0,1,2,\ldots\,;\,m=-l,\ldots,l)$. Defining the spin $s$ spherical harmonics as
\begin{equation}
	\Preind_{s}{Y}_{l,m}=\begin{cases}
	\sqrt{\dfrac{(l-s)!}{2(l+s)!}}\eth^s Y_{l,m}\qquad\qquad(0\leq s\leq l)\\
	(-1)^s\sqrt{\dfrac{(l+s)!}{2(l-s)!}}\xbar\eth^{-s} Y_{l,m}\quad(-l\leq s\leq 0)
\end{cases} \ ,
\end{equation}
the following relations can be deduced:
\begin{equation}
\label{zero}
\begin{gathered}
	\int\td z \td\bz\,\gamma_{z\bz}\, \Preind_{s}{Y}_{l,m}\,\eth^{l-s+1}\eta =0 \ ,\qquad
	\int \td z \td\bz\,\gamma_{z\bz}\,\Preind_{s}{\bar{Y}}_{l,m}  \,\xbar\eth^{l-s+1}\zeta=0 \ ,\\
	\xbar\eth\eth\Preind_{s}{Y}_{l,m}=-\half(l-s)(l+s+1)\Preind_{s}{Y}_{l,m} \ , \qquad
	\int\td z \td\bz\,\gamma_{z\bz}\, A\eth B=-\int\td z \td\bz\,\gamma_{z\bz}\, B\eth A \ .
\end{gathered}
\end{equation}
where $\eta$ and $\zeta$ have spin weight $-l-1$ and $l+1$ respectively, while $A\eth B$ has no spin weight. These properties allow for a compact definition of conserved charges below.

Following \cite{Newman:1968uj}, the Maxwell-tensors are replaced by three complex scalars:
\begin{equation}
\label{phi012}
	\phi_0=F_{r\bz}\dfrac{\ga^{-\half}}{r} \ ,\quad
	\phi_1=\half(F_{ru}+F_{z\bz}\dfrac{\gai}{r^2}) \ ,\quad
	\phi_2=\dfrac{\ga^{-\half}}{r}(F_{zu}-\half F_{zr}) \ ,
\end{equation}
With these quantities, the vacuum Maxwell equations can be organized in the Newman-Penrose formalism as
\begin{equation}
\begin{aligned}
	\p_r(r^2\phi_1)&=r \xbar\eth\phi_0 \,	&	(\p_u-\half \p_r -\dfrac{1}{2 r})\phi_0&=\dfrac{\eth\phi_1}{r} \ , \\
	\p_r(r\phi_2)&=r \xbar\eth\phi_1 \ ,	&	(\p_u-\half \p_r - \dfrac{1}{r})\phi_1&=\dfrac{\eth\phi_2}{r} \ .
\end{aligned}
\end{equation}
Assuming the ansatz~\eqref{Azz}, the solution is
\begin{equation}
\label{solM}
\begin{aligned}
	\phi_0&= \sum\limits_{n=0}^\infty \frac{\phi^n_0(u,z,\bz)}{r^{n+3}} \ , &
	\p_u\phi_1^0&=\eth\phi_2^0 \ , \\
	\phi_1&=\dfrac{\phi_1^0}{r^2} - \sum\limits_{n=0}^\infty \frac{\xbar\eth\phi^n_0(u,z,\bz)}{(n+1)r^{n+3}} \ , &
	\p_u\phi_0^0&=\eth\phi_1^0 \ , \\
	\phi_2&=\dfrac{\phi_2^0}{r}-\dfrac{\xbar\eth\phi_1^0}{r^2} +
	\sum\limits_{n=0}^\infty \frac{\xbar\eth^2\phi^n_0(u,z,\bz)}{(n+1)(n+2)r^{n+3}} \ ,\quad &
	\p_u\phi_0^{n+1}&=-\dfrac{n+2}{2}\phi_0^n - \dfrac{\eth\xbar\eth\phi_0^n}{n+1} \ .
\end{aligned}
\end{equation}
Confirming the analysis that we did in Section~\ref{sos}, we see that, in the end, $\phi_2^0(u,z,\bz)$ is the news function of this system, associated to electromagnetic radiations. To be more specific, the concrete translation to the expressions in the main text is
\begin{equation}
	\phi_0^n=-(n+1)\ga^{-\half}A^{n+1}_{\bz} \  , \quad
	\phi^0_1=-\half[A_u^0-\gai(\p_z A_{\bz}^0-\p_{\bz} A_z^0)] \ ,\quad
	\phi^0_2=-\ga^{-\half}\p_u A_z^0 \ ,
\end{equation}
where it is understood that $n$ is a non-negative integer. Notice that $\phi_i^0$ ($i=0,1,2$) only involve the boundary fields $A^0_u$, $A^0_{z(\bz)}$ and $A^1_{z(\bz)}$.
The physical meaning of $\phi_1^0$ and $\phi_0^n$ was already identified in \cite{Janis:1965tx}. The real and imaginary parts of $\phi_1^0$ are the electric and magnetic charges respectively. Similarly, real and imaginary parts of $\xbar\eth\phi_0^n$ are the electric and magnetic multipoles. In particular, the dipole corresponds to $\xbar\eth\phi_0^0$. Moreover, the spin weights of $\phi_2^0$, $\phi_1^0$ and $\phi_0^n$ are respectively $-1,0,1$. Then, using~\eqref{solM} and~\eqref{zero}, we can construct conserved quantities
\bea
&&\p_u\int\td z \td\bz\,\gamma_{z\bz}\, \Preind_{0}{Y}_{0,0}\,\phi_1^0=0 \ ,\\
&&\p_u\int\td z \td\bz\,\gamma_{z\bz}\, \Preind_{0}{Y}_{l,m}\,\xbar\eth\phi_0^l =0 \ .\quad (l=0,1,2,\ldots) \ .
\eea
Notice that these conserved quantities all follow from the different orders of $\phi_1$.
When $l=0$ we can see that $\int\td z \td\bz\, \Preind_{0}{Y}_{0,0}\,\xbar\eth\phi_0^0$ is automatically vanishing. This means that none of the Newman-Penrose charges correspond to dipole charge. Therefore, to us it does not seem possible to interpret~\eqref{Q1} as a generalization of dipole charge.
\todo{can we interpret as generalization of other NP charge?}

As a final remark, let us on comment on the fact that the Newman-Penrose construction for Einstein gravity is very similar. But instead of the three boundary fields that we had in~\eqref{phi012} (\textit{i.e.} the ``news'' $\phi_2^0$, plus $\phi_1^0$, and $\phi_0^0$), there we will have four boundary fields, namely the Bondi ``news'' and other three ($\Psi^0_2$, $\Psi^0_1$ and $\Psi^0_0$ in the standard Newman-Penrose notation) coming from the Weyl tensor. This clearly hints at the fact that in Einstein gravity there is an extra sub-sub-leading order in the soft theorem~\cite{Cachazo:2014fwa}.

\bibliography{amplitudesrefs,asymptrefs,local}

\providecommand{\href}[2]{#2}\begingroup\raggedright\begin{thebibliography}{10}

\bibitem{Strominger:2013lka}
A.~Strominger, ``{Asymptotic Symmetries of Yang-Mills Theory},''
  \href{http://dx.doi.org/10.1007/JHEP07(2014)151}{{\em JHEP} {\bfseries 07}
  (2014) 151},
\href{http://arxiv.org/abs/1308.0589}{{\ttfamily arXiv:1308.0589 [hep-th]}}.
%%CITATION = ARXIV:1308.0589;%%.

\bibitem{Strominger:2013jfa}
A.~Strominger, ``{On BMS Invariance of Gravitational Scattering},''
  \href{http://dx.doi.org/10.1007/JHEP07(2014)152}{{\em JHEP} {\bfseries 07}
  (2014) 152},
\href{http://arxiv.org/abs/1312.2229}{{\ttfamily arXiv:1312.2229 [hep-th]}}.
%%CITATION = ARXIV:1312.2229;%%.

\bibitem{Hawking:2016msc}
S.~W. Hawking, M.~J. Perry, and A.~Strominger, ``{Soft Hair on Black Holes},''
  \href{http://dx.doi.org/10.1103/PhysRevLett.116.231301}{{\em Phys. Rev.
  Lett.} {\bfseries 116} no.~23, (2016) 231301},
\href{http://arxiv.org/abs/1601.00921}{{\ttfamily arXiv:1601.00921 [hep-th]}}.
%%CITATION = ARXIV:1601.00921;%%.

\bibitem{Kapec:2016jld}
D.~Kapec, P.~Mitra, A.-M. Raclariu, and A.~Strominger, ``{A 2D Stress Tensor
  for 4D Gravity},''
\href{http://arxiv.org/abs/1609.00282}{{\ttfamily arXiv:1609.00282 [hep-th]}}.
%%CITATION = ARXIV:1609.00282;%%.

\bibitem{Cheung:2016iub}
C.~Cheung, A.~de~la Fuente, and R.~Sundrum, ``{4D Scattering Amplitudes and
  Asymptotic Symmetries from 2D CFT},''
\href{http://arxiv.org/abs/1609.00732}{{\ttfamily arXiv:1609.00732 [hep-th]}}.
%%CITATION = ARXIV:1609.00732;%%.

\bibitem{Bondi:1962px}
H.~Bondi, M.~G.~J. van~der Burg, and A.~W.~K. Metzner, ``{Gravitational waves
  in general relativity. 7. Waves from axisymmetric isolated systems},''
\href{http://dx.doi.org/10.1098/rspa.1962.0161}{{\em Proc. Roy. Soc. Lond.}
  {\bfseries A269} (1962) 21--52}.
%%CITATION = PRSLA,A269,21;%%.

\bibitem{Sachs:1962wk}
R.~K. Sachs, ``{Gravitational waves in general relativity. 8. Waves in
  asymptotically flat space-times},''
\href{http://dx.doi.org/10.1098/rspa.1962.0206}{{\em Proc. Roy. Soc. Lond.}
  {\bfseries A270} (1962) 103--126}.
%%CITATION = PRSLA,A270,103;%%.

\bibitem{Barnich:2009se}
G.~Barnich and C.~Troessaert, ``{Symmetries of asymptotically flat 4
  dimensional spacetimes at null infinity revisited},''
  \href{http://dx.doi.org/10.1103/PhysRevLett.105.111103}{{\em Phys. Rev.
  Lett.} {\bfseries 105} (2010) 111103},
\href{http://arxiv.org/abs/0909.2617}{{\ttfamily arXiv:0909.2617 [gr-qc]}}.
%%CITATION = ARXIV:0909.2617;%%.

\bibitem{Barnich:2010eb}
G.~Barnich and C.~Troessaert, ``{Aspects of the BMS/CFT correspondence},''
  \href{http://dx.doi.org/10.1007/JHEP05(2010)062}{{\em JHEP} {\bfseries 05}
  (2010) 062},
\href{http://arxiv.org/abs/1001.1541}{{\ttfamily arXiv:1001.1541 [hep-th]}}.
%%CITATION = ARXIV:1001.1541;%%.

\bibitem{Barnich:2011ct}
G.~Barnich and C.~Troessaert, ``{Supertranslations call for superrotations},''
  {\em PoS} (2010) 010, \href{http://arxiv.org/abs/1102.4632}{{\ttfamily
  arXiv:1102.4632 [gr-qc]}}.
[Ann. U. Craiova Phys.21,S11(2011)].
%%CITATION = ARXIV:1102.4632;%%.

\bibitem{Barnich:2011mi}
G.~Barnich and C.~Troessaert, ``{BMS charge algebra},''
  \href{http://dx.doi.org/10.1007/JHEP12(2011)105}{{\em JHEP} {\bfseries 12}
  (2011) 105},
\href{http://arxiv.org/abs/1106.0213}{{\ttfamily arXiv:1106.0213 [hep-th]}}.
%%CITATION = ARXIV:1106.0213;%%.

\bibitem{Cachazo:2014fwa}
F.~Cachazo and A.~Strominger, ``{Evidence for a New Soft Graviton Theorem},''
\href{http://arxiv.org/abs/1404.4091}{{\ttfamily arXiv:1404.4091 [hep-th]}}.
%%CITATION = ARXIV:1404.4091;%%.

\bibitem{Gross:1968in}
D.~J. Gross and R.~Jackiw, ``{Low-Energy Theorem for Graviton Scattering},''
\href{http://dx.doi.org/10.1103/PhysRev.166.1287}{{\em Phys. Rev.} {\bfseries
  166} (1968) 1287--1292}.
%%CITATION = PHRVA,166,1287;%%.

\bibitem{White:2011yy}
C.~D. White, ``{Factorization Properties of Soft Graviton Amplitudes},''
  \href{http://dx.doi.org/10.1007/JHEP05(2011)060}{{\em JHEP} {\bfseries 05}
  (2011) 060},
\href{http://arxiv.org/abs/1103.2981}{{\ttfamily arXiv:1103.2981 [hep-th]}}.
%%CITATION = ARXIV:1103.2981;%%.

\bibitem{Casali:2014xpa}
E.~Casali, ``{Soft sub-leading divergences in Yang-Mills amplitudes},''
  \href{http://dx.doi.org/10.1007/JHEP08(2014)077}{{\em JHEP} {\bfseries 08}
  (2014) 077},
\href{http://arxiv.org/abs/1404.5551}{{\ttfamily arXiv:1404.5551 [hep-th]}}.
%%CITATION = ARXIV:1404.5551;%%.

\bibitem{Laenen:2008gt}
E.~Laenen, G.~Stavenga, and C.~D. White, ``{Path integral approach to eikonal
  and next-to-eikonal exponentiation},''
  \href{http://dx.doi.org/10.1088/1126-6708/2009/03/054}{{\em JHEP} {\bfseries
  03} (2009) 054},
\href{http://arxiv.org/abs/0811.2067}{{\ttfamily arXiv:0811.2067 [hep-ph]}}.
%%CITATION = ARXIV:0811.2067;%%.

\bibitem{Laenen:2010uz}
E.~Laenen, L.~Magnea, G.~Stavenga, and C.~D. White, ``{Next-to-eikonal
  corrections to soft gluon radiation: a diagrammatic approach},''
  \href{http://dx.doi.org/10.1007/JHEP01(2011)141}{{\em JHEP} {\bfseries 01}
  (2011) 141},
\href{http://arxiv.org/abs/1010.1860}{{\ttfamily arXiv:1010.1860 [hep-ph]}}.
%%CITATION = ARXIV:1010.1860;%%.

\bibitem{White:2014qia}
C.~D. White, ``{Diagrammatic insights into next-to-soft corrections},''
  \href{http://dx.doi.org/10.1016/j.physletb.2014.08.041}{{\em Phys. Lett.}
  {\bfseries B737} (2014) 216--222},
\href{http://arxiv.org/abs/1406.7184}{{\ttfamily arXiv:1406.7184 [hep-th]}}.
%%CITATION = ARXIV:1406.7184;%%.

\bibitem{Low:1954kd}
F.~E. Low, ``{Scattering of light of very low frequency by systems of spin
  1/2},''
\href{http://dx.doi.org/10.1103/PhysRev.96.1428}{{\em Phys. Rev.} {\bfseries
  96} (1954) 1428--1432}.
%%CITATION = PHRVA,96,1428;%%.

\bibitem{GellMann:1954kc}
M.~Gell-Mann and M.~L. Goldberger, ``{Scattering of low-energy photons by
  particles of spin 1/2},''
\href{http://dx.doi.org/10.1103/PhysRev.96.1433}{{\em Phys. Rev.} {\bfseries
  96} (1954) 1433--1438}.
%%CITATION = PHRVA,96,1433;%%.

\bibitem{Low:1958sn}
F.~E. Low, ``{Bremsstrahlung of very low-energy quanta in elementary particle
  collisions},''
\href{http://dx.doi.org/10.1103/PhysRev.110.974}{{\em Phys. Rev.} {\bfseries
  110} (1958) 974--977}.
%%CITATION = PHRVA,110,974;%%.

\bibitem{Bianchi:2014gla}
M.~Bianchi, S.~He, Y.-t. Huang, and C.~Wen, ``{More on Soft Theorems: Trees,
  Loops and Strings},''
  \href{http://dx.doi.org/10.1103/PhysRevD.92.065022}{{\em Phys. Rev.}
  {\bfseries D92} no.~6, (2015) 065022},
\href{http://arxiv.org/abs/1406.5155}{{\ttfamily arXiv:1406.5155 [hep-th]}}.
%%CITATION = ARXIV:1406.5155;%%.

\bibitem{Burnett:1967km}
T.~H. Burnett and N.~M. Kroll, ``{Extension of the low soft photon theorem},''
\href{http://dx.doi.org/10.1103/PhysRevLett.20.86}{{\em Phys. Rev. Lett.}
  {\bfseries 20} (1968) 86}.
%%CITATION = PRLTA,20,86;%%.

\bibitem{Bell:1969yw}
J.~S. Bell and R.~Van~Royen, ``{On the low-burnett-kroll theorem for
  soft-photon emission},''
\href{http://dx.doi.org/10.1007/BF02823297}{{\em Nuovo Cim.} {\bfseries A60}
  (1969) 62--68}.
%%CITATION = NUCIA,A60,62;%%.

\bibitem{Weinberg:1965nx}
S.~Weinberg, ``{Infrared photons and gravitons},''
\href{http://dx.doi.org/10.1103/PhysRev.140.B516}{{\em Phys. Rev.} {\bfseries
  140} (1965) B516--B524}.
%%CITATION = PHRVA,140,B516;%%.

\bibitem{DelDuca:1990gz}
V.~Del~Duca, ``{High-energy Bremsstrahlung Theorems for Soft Photons},''
\href{http://dx.doi.org/10.1016/0550-3213(90)90392-Q}{{\em Nucl. Phys.}
  {\bfseries B345} (1990) 369--388}.
%%CITATION = NUPHA,B345,369;%%.

\bibitem{Larkoski:2014bxa}
A.~J. Larkoski, D.~Neill, and I.~W. Stewart, ``{Soft Theorems from Effective
  Field Theory},'' \href{http://dx.doi.org/10.1007/JHEP06(2015)077}{{\em JHEP}
  {\bfseries 06} (2015) 077},
\href{http://arxiv.org/abs/1412.3108}{{\ttfamily arXiv:1412.3108 [hep-th]}}.
%%CITATION = ARXIV:1412.3108;%%.

\bibitem{Kulish:1970ut}
P.~P. Kulish and L.~D. Faddeev, ``{Asymptotic conditions and infrared
  divergences in quantum electrodynamics},''
  \href{http://dx.doi.org/10.1007/BF01066485}{{\em Theor. Math. Phys.}
  {\bfseries 4} (1970) 745}.
[Teor. Mat. Fiz.4,153(1970)].
%%CITATION = TMPHA,4,745;%%.

\bibitem{Gomez:2016hxz}
C.~Gomez and M.~Panchenko, ``{Asymptotic dynamics, large gauge transformations
  and infrared symmetries},''
\href{http://arxiv.org/abs/1608.05630}{{\ttfamily arXiv:1608.05630 [hep-th]}}.
%%CITATION = ARXIV:1608.05630;%%.

\bibitem{He:2014cra}
T.~He, P.~Mitra, A.~P. Porfyriadis, and A.~Strominger, ``{New Symmetries of
  Massless QED},'' \href{http://dx.doi.org/10.1007/JHEP10(2014)112}{{\em JHEP}
  {\bfseries 10} (2014) 112},
\href{http://arxiv.org/abs/1407.3789}{{\ttfamily arXiv:1407.3789 [hep-th]}}.
%%CITATION = ARXIV:1407.3789;%%.

\bibitem{Kapec:2014zla}
D.~Kapec, V.~Lysov, and A.~Strominger, ``{Asymptotic Symmetries of Massless QED
  in Even Dimensions},''
\href{http://arxiv.org/abs/1412.2763}{{\ttfamily arXiv:1412.2763 [hep-th]}}.
%%CITATION = ARXIV:1412.2763;%%.

\bibitem{Mohd:2014oja}
A.~Mohd, ``{A note on asymptotic symmetries and soft-photon theorem},''
  \href{http://dx.doi.org/10.1007/JHEP02(2015)060}{{\em JHEP} {\bfseries 02}
  (2015) 060},
\href{http://arxiv.org/abs/1412.5365}{{\ttfamily arXiv:1412.5365 [hep-th]}}.
%%CITATION = ARXIV:1412.5365;%%.

\bibitem{Kapec:2015ena}
D.~Kapec, M.~Pate, and A.~Strominger, ``{New Symmetries of QED},''
\href{http://arxiv.org/abs/1506.02906}{{\ttfamily arXiv:1506.02906 [hep-th]}}.
%%CITATION = ARXIV:1506.02906;%%.

\bibitem{Campiglia:2015qka}
M.~Campiglia and A.~Laddha, ``{Asymptotic symmetries of QED and Weinberg’s
  soft photon theorem},'' \href{http://dx.doi.org/10.1007/JHEP07(2015)115}{{\em
  JHEP} {\bfseries 07} (2015) 115},
\href{http://arxiv.org/abs/1505.05346}{{\ttfamily arXiv:1505.05346 [hep-th]}}.
%%CITATION = ARXIV:1505.05346;%%.

\bibitem{Strominger:2015bla}
A.~Strominger, ``{Magnetic Corrections to the Soft Photon Theorem},''
  \href{http://dx.doi.org/10.1103/PhysRevLett.116.031602}{{\em Phys. Rev.
  Lett.} {\bfseries 116} no.~3, (2016) 031602},
\href{http://arxiv.org/abs/1509.00543}{{\ttfamily arXiv:1509.00543 [hep-th]}}.
%%CITATION = ARXIV:1509.00543;%%.

\bibitem{Lysov:2014csa}
V.~Lysov, S.~Pasterski, and A.~Strominger, ``{Low’s Subleading Soft Theorem
  as a Symmetry of QED},''
  \href{http://dx.doi.org/10.1103/PhysRevLett.113.111601}{{\em Phys. Rev.
  Lett.} {\bfseries 113} no.~11, (2014) 111601},
\href{http://arxiv.org/abs/1407.3814}{{\ttfamily arXiv:1407.3814 [hep-th]}}.
%%CITATION = ARXIV:1407.3814;%%.

\bibitem{Broedel:2014fsa}
J.~Broedel, M.~de~Leeuw, J.~Plefka, and M.~Rosso, ``{Constraining subleading
  soft gluon and graviton theorems},''
  \href{http://dx.doi.org/10.1103/PhysRevD.90.065024}{{\em Phys. Rev.}
  {\bfseries D90} no.~6, (2014) 065024},
\href{http://arxiv.org/abs/1406.6574}{{\ttfamily arXiv:1406.6574 [hep-th]}}.
%%CITATION = ARXIV:1406.6574;%%.

\bibitem{Bern:2014vva}
Z.~Bern, S.~Davies, P.~Di~Vecchia, and J.~Nohle, ``{Low-Energy Behavior of
  Gluons and Gravitons from Gauge Invariance},''
  \href{http://dx.doi.org/10.1103/PhysRevD.90.084035}{{\em Phys. Rev.}
  {\bfseries D90} no.~8, (2014) 084035},
\href{http://arxiv.org/abs/1406.6987}{{\ttfamily arXiv:1406.6987 [hep-th]}}.
%%CITATION = ARXIV:1406.6987;%%.

\bibitem{Janis:1965tx}
A.~I. Janis and E.~T. Newman, ``{Structure of Gravitational Sources},''
\href{http://dx.doi.org/10.1063/1.1704349}{{\em J. Math. Phys.} {\bfseries 6}
  (1965) 902--914}.
%%CITATION = JMAPA,6,902;%%.

\bibitem{Newman:1968uj}
E.~T. Newman and R.~Penrose, ``{New conservation laws for zero rest-mass fields
  in asymptotically flat space-time},''
\href{http://dx.doi.org/10.1098/rspa.1968.0112}{{\em Proc. Roy. Soc. Lond.}
  {\bfseries A305} (1968) 175--204}.
%%CITATION = PRSLA,A305,175;%%.

\bibitem{Barnich:2015jua}
G.~Barnich, P.-H. Lambert, and P.-J. Mao, ``{Three-dimensional asymptotically
  flat Einstein–Maxwell theory},''
  \href{http://dx.doi.org/10.1088/0264-9381/32/24/245001}{{\em Class. Quant.
  Grav.} {\bfseries 32} no.~24, (2015) 245001},
\href{http://arxiv.org/abs/1503.00856}{{\ttfamily arXiv:1503.00856 [gr-qc]}}.
%%CITATION = ARXIV:1503.00856;%%.

\bibitem{Barnich:2001jy}
G.~Barnich and F.~Brandt, ``{Covariant theory of asymptotic symmetries,
  conservation laws and central charges},''
  \href{http://dx.doi.org/10.1016/S0550-3213(02)00251-1}{{\em Nucl. Phys.}
  {\bfseries B633} (2002) 3--82},
\href{http://arxiv.org/abs/hep-th/0111246}{{\ttfamily arXiv:hep-th/0111246
  [hep-th]}}.
%%CITATION = HEP-TH/0111246;%%.

\bibitem{Ashtekar:1987tt}
A.~Ashtekar, {\em {ASYMPTOTIC QUANTIZATION: BASED ON 1984 NAPLES LECTURES}}.
\newblock
1987.
\newblock
%%CITATION = INSPIRE-256903;%%.

\bibitem{DiVecchia:2015jaq}
P.~Di~Vecchia, R.~Marotta, M.~Mojaza, and J.~Nohle, ``{New soft theorems for
  the gravity dilaton and the Nambu-Goldstone dilaton at subsubleading
  order},'' \href{http://dx.doi.org/10.1103/PhysRevD.93.085015}{{\em Phys.
  Rev.} {\bfseries D93} no.~8, (2016) 085015},
\href{http://arxiv.org/abs/1512.03316}{{\ttfamily arXiv:1512.03316 [hep-th]}}.
%%CITATION = ARXIV:1512.03316;%%.

\bibitem{Kapec:2014opa}
D.~Kapec, V.~Lysov, S.~Pasterski, and A.~Strominger, ``{Semiclassical Virasoro
  symmetry of the quantum gravity $ \mathcal{S}$-matrix},''
  \href{http://dx.doi.org/10.1007/JHEP08(2014)058}{{\em JHEP} {\bfseries 08}
  (2014) 058},
\href{http://arxiv.org/abs/1406.3312}{{\ttfamily arXiv:1406.3312 [hep-th]}}.
%%CITATION = ARXIV:1406.3312;%%.

\bibitem{Campiglia:2014yka}
M.~Campiglia and A.~Laddha, ``{Asymptotic symmetries and subleading soft
  graviton theorem},'' \href{http://dx.doi.org/10.1103/PhysRevD.90.124028}{{\em
  Phys. Rev.} {\bfseries D90} no.~12, (2014) 124028},
\href{http://arxiv.org/abs/1408.2228}{{\ttfamily arXiv:1408.2228 [hep-th]}}.
%%CITATION = ARXIV:1408.2228;%%.

\bibitem{Campiglia:2015yka}
M.~Campiglia and A.~Laddha, ``{New symmetries for the Gravitational
  S-matrix},'' \href{http://dx.doi.org/10.1007/JHEP04(2015)076}{{\em JHEP}
  {\bfseries 04} (2015) 076},
\href{http://arxiv.org/abs/1502.02318}{{\ttfamily arXiv:1502.02318 [hep-th]}}.
%%CITATION = ARXIV:1502.02318;%%.

\bibitem{2come}
E.~Conde and P.~Mao {\em {work in progress}} .

\end{thebibliography}\endgroup
\bibliographystyle{utphys}
\end{document}